\documentclass[twocolumn]{aastex63}
\usepackage{soul}
\usepackage{xcolor}
\usepackage{amsmath}
\usepackage{textcomp}
\usepackage{multirow}
\usepackage{graphicx}
\usepackage{booktabs}
\usepackage{tabularx}
\usepackage{rotating}
\usepackage{makecell}
\usepackage{mathrsfs}
\usepackage{mathtools}
\usepackage{pdfcomment}
\usepackage[T1]{fontenc}
\usepackage[utf8]{inputenc}

\newcommand{\msun}{M_\odot}
\newcommand{\mstar}{M_\star}
\newcommand{\mearth}{M_\oplus}
\newcommand{\cs}{c_{\text{s}}}
\newcommand{\rhog}{\rho_{\text{g}}}
\newcommand{\rhod}{\rho_{\text{d}}}
\newcommand{\I}{\textit{\textbf{I}}}
\newcommand{\V}{\textit{\textbf{V}}}
\newcommand{\W}{\textit{\textbf{W}}}
\newcommand{\sigmag}{\Sigma_{\text{g}}}

\defcitealias{drazkowska_planetesimal_2017}{Dr{\k{a}}{\.z}kowska \& Alibert 2017}

\begin{document}

\title{Puffed up Edges of Planet-opened Gaps in Protoplanetary Disks. I. hydrodynamic simulations}

\author[0000-0002-0605-4961]{Jiaqing Bi}
\affil{Department of Physics \& Astronomy, University of Victoria, 3800 Finnerty Road, Victoria, BC V8P 5C2, Canada}
\affil{Academia Sinica Institute of Astronomy \& Astrophysics, No. 1, Sec. 4, Roosevelt Rd, Taipei 10617, Taiwan}
\email{bijiaqing@uvic.ca}

\author[0000-0002-8597-4386]{Min-Kai Lin}
\affil{Academia Sinica Institute of Astronomy \& Astrophysics, No. 1, Sec. 4, Roosevelt Rd, Taipei 10617, Taiwan}
\affil{Physics Division, National Center for Theoretical Sciences, Taipei 10617, Taiwan}

\author[0000-0001-9290-7846]{Ruobing Dong}
\affil{Department of Physics \& Astronomy, University of Victoria, 3800 Finnerty Road, Victoria, BC V8P 5C2, Canada}


\begin{abstract}

Dust gaps and rings appear ubiquitous in bright protoplanetary disks. Disk-planet interaction with dust-trapping at the edges of planet-induced gaps is one plausible explanation. However, the sharpness of some observed dust rings indicate that sub-mm-sized dust grains have settled to a thin layer in some systems. We test whether or not such dust around gas gaps opened by planets can remain settled by performing three-dimensional, dust-plus-gas simulations of protoplanetary disks with an embedded planet. We find planets massive enough to open gas gaps stir small, sub-mm-sized dust grains to high disk elevations at the gap edges, where the dust scale-height can reach $\sim 70\%$ of the gas scale-height. We attribute this dust `puff-up' to the planet-induced meridional gas flows previously identified by \citeauthor{fung_gap_2016} and others. We thus emphasize the importance of explicit 3D simulations to obtain the vertical distribution of sub-mm-sized grains around gas gaps opened by massive planets. We caution that the gas-gap-opening planet interpretation of well-defined dust rings is only self-consistent with large grains exceeding mm in size.  

\end{abstract}

\keywords{protoplanetary disks -- planet formation -- disk-planet interaction -- hydrodynamics -- methods: numerical}


\section{Introduction}

One of the most exciting developments in the field of planet formation is the direct observation of detailed sub-structures in protoplanetary disks (PPDs). The Atacama Large Millimeter/submillimeter Array (ALMA) has shown that many of the largest PPDs contain dust gaps and rings \citep{andrews_disk_2018, long_gaps_2018, van_der_marel_protoplanetary_2019}, while a smaller, but non-negligible fraction contain asymmetries such as lopsided dust clumps and spiral arms \citep[e.g.,][]{van_der_marel_major_2013, dong_eccentric_2018}. 

The planet interpretation of dust rings has become an attractive scenario as pressure bumps naturally arise from the gap-opening process by massive planets \citep{lin_tidal_1993}. Solids can then be trapped at the two gap edges on either side of the planet \citep{paardekooper_planets_2004, paardekooper_dust_2006, rosotti_minimum_2016, dipierro_opening_2017, weber_predicting_2019, meru_is_2019, yang_morphological_2019}. It is also possible for a single planet to open additional gaps away from its orbital radius \citep{bae_formation_2017} and thus produce more than two dust rings \citep{dong_multiple_2017, dong_multiple_2018}. An accurate model of planet-induced dust rings can provide an indirect method to detect and characterize planets (as well as disk properties) during their formation \citep{lodato_newborn_2019, zhang_disk_2018}. 

Dust rings associated with planet gaps are often modeled assuming a two-dimensional (2D), razor-thin PPD. Such models either represent a vertically-integrated system, or focus on conditions close to the disk midplane. This is useful, and often necessary, for reducing the computational cost to cover the large parameter space intrinsic to disk-planet interaction \citep[e.g.,][]{zhang_effects_2020} and/or to perform high resolution simulations \citep[e.g.,][]{hsieh_migrating_2020, mcnally_migrating_2019}. Indeed, this approximation allows one to construct empirical models of planet gaps based on large sets of simulation data \citep{kanagawa_mass_2016, auddy_machine_2020}.

However, real PPDs are three-dimensional (3D). An important effect in 3D is the settling of solids to the disk midplane due to the gravity from the central star in the vertical direction \citep{dubrulle_dust_1995, youdin_particle_2007, laibe_settling_2020}. Traditionally, dust settling is assumed to be balanced by turbulent diffusion, resulting in a finite dust layer thickness. This, in turn, directly affects the appearance of dust rings. For example, the sharp, well-defined dust rings observed in the disk around HL Tau suggest that dust grains are well-settled \citep{pinte_dust_2016}.

On the other hand, it is not clear if flattened dust layers are consistent with the interpretation of observed dust rings being produced by giant planets. This is because such planets can induce complex, 3D gas flows \citep{morbidelli_meridional_2014, szulagyi_accretion_2014, fung_gap_2016, bae_spiral_2016, dong_observational_2019, teague_meridional_2019}. Specifically, \cite{fung_gap_2016} showed that gap-opening planets induce large-scale meridional circulations around gap edges, which originate from the differential vertical dependence of planet and viscous torques. If dust grains cannot settle against these meridional circulations, then planet gaps may not explain sharp, well-defined dust rings observed in real PPDs.

In fact, already in a pioneering (pre-ALMA) study, \cite{fouchet_effect_2007} carried out 3D Smoothed Particle Hydrodynamic simulations of planets embedded in dusty disks. Although they did not focus on the issue of meridional flows and dust settling, their simulations indicate that while large meter-sized bodies settle, smaller, cm-sized grains have significantly larger scale-heights. This is an alarming result because grain sizes in the ALMA-observed disks may be even smaller, perhaps only up to 0.1--1 mm \citep{kataoka_grain_2016, kataoka_evidence_2017, liu_anomalously_2019, zhu_one_2019}, and are thus more easily stirred by gas motions. 

Are sharp, flattened dust rings in sub-mm-sized grains observed in PPDs compatible with gas gaps opened by planets? To address this issue, we perform grid-based hydrodynamic simulations of 3D, dusty PPDs with embedded planets. We find gas-gap-opening planets efficiently stir sub-mm-sized dust grains to high elevations at gap edges. We attribute this to the planet-induced meridional gas flows identified by \cite{fung_gap_2016}. Our results suggest that 3D models of dusty gaps can be used to constrain the minimum grain size, the planet mass, or both, associated with well-defined, planet-induced dust rings in PPDs. 

This paper is organized as follows. We first describe the disk-planet system of interest and its numerical modeling in Section \ref{sec:model}. We present results in Section \ref{sec:result}, starting with a fiducial case, followed by a brief parameter survey. Here, we demonstrate that the `puff-up' of the dust layer primarily depends on grain size and, to a lesser degree, the planet mass. We discuss the implications of our results in Section \ref{sec:discuss} and conclude in Section \ref{sec:summary}. 


\section{Disk-planet Model} \label{sec:model}

We consider a 3D protoplanetary disk composed of gas and dust with an embedded planet of mass $M_{\text{p}}$ around a central star of mass $\mstar$. We neglect disk self-gravity, magnetic fields, planet orbital migration, and planet accretion. Hereafter, \{$r, \phi, \theta$\} denote spherical radius, azimuth, and polar angle, while \{$R, \phi, Z$\} denote cylindrical radius, azimuth, and height. Both coordinates are centered on the star. We use the subscript `ref' to denote evaluations in the midplane at $R = R_\text{ref}$, where $R_\text{ref}$ is a reference radius. We use the subscript `0' to denote initial values.

The volume density, pressure, and velocity of gas are denoted by ($\rhog, P, \V\,$). We assume a time-independent, vertically isothermal, axisymmetric gas temperature profile 
\begin{equation}
T(R) = T_\text{ref} \left(\frac{R}{R_\text{ref}}\right)^{-q},
\end{equation}
where $q$ is a constant power-law index. The corresponding sound speed is
\begin{equation}
\cs(R) = c_{\text{s,ref}}\left(\frac{R}{R_\text{ref}}\right)^{-q/2}, 
\end{equation}
and our isothermal equation of state is
\begin{equation}
P = \rhog \cs^2.
\end{equation}
The pressure scale-height of gas is defined as 
\begin{equation}
H_{\text{g}} = \frac{\cs}{\Omega_{\text{K}}},
\end{equation}
where $\Omega_{\text{K}}(R) = \sqrt{G\mstar/R^3}$ is the Keplerian angular velocity and $G$ is the gravitational constant. We assume a nonflared gas disk with a constant aspect ratio $h_{\text{g}} \equiv H_{\text{g}}/R = 0.05$, corresponding to $q = 1$. 

We consider a single species of dust modeled as a pressureless fluid with density and velocity ($\rhod, \W\,$). Dust-gas coupling is parameterized by the Stokes number 
\begin{equation}
\text{St} = \tau_{\text{s}}\Omega_{\text{K}}, 
\end{equation}
where $\tau_\mathrm{s}$ is the particle stopping time characterizing the frictional drag force between gas and dust. We consider dust tightly (but imperfectly) coupled to the gas with $\text{St}\ll 1$ \citep{jacquet_linear_2011}. 

We assume the dust grains are in the Epstein regime with fixed grain size $s$ and internal density $\rho_{\bullet}$. The particle stopping time is $\tau_\text{s} = \rho_{\bullet}s/\rhog\cs$ \citep{weidenschilling_aerodynamics_1977}. In practice, we adopt the prescription 
\begin{equation} \label{eq:tstop}
\tau_{\text{s}} = 
\frac{\rho_\mathrm{g0,ref}}{\rhog}
\frac{c_{\text{s,ref}}}{\cs}
\frac{\text{St}_\text{0,ref}}{\Omega_{\text{K,ref}}},
\end{equation}
and choose a reference Stokes number $\text{St}_\text{0,ref} = 10^{-3}$ to represent 0.1-mm-sized grains with $\rho_{\bullet}$ = 1.5 g cm\textsuperscript{-3} at $\sim 45$ astronomical units (au) in young PPDs, such as the HL Tau disk\footnote{Assuming the total disk mass being 0.2 $\msun$ \citep{booth_13c17o_2020}, the outer disk radius being 150 au, and the surface density power-law index being -1.5.}.


\subsection{Basic equations}

The PPD described above is governed by the usual hydrodynamic equations for gas and dust,  
\begin{align} 
&\frac{\partial \rhog}{\partial t} + \nabla \cdot (\rhog \V\,) = 0, \\
&\frac{\partial \V}{\partial t} + \V \cdot \nabla \V = - \frac{1}{\rhog} \nabla P
 - \nabla \Phi + \frac{\epsilon}{\tau_{\text{s}}}(\W - \V) + \frac{1}{\rhog} \nabla \cdot \mathcal{T},\\
&\frac{\partial \rhod}{\partial t} + \nabla \cdot (\rhod \W\,) = 0, \\ 
&\frac{\partial \W}{\partial t} + \W \cdot \nabla \W =  - \nabla \Phi - \frac{1}{\tau_{\text{s}}}(\W - \V).
\end{align}
$\Phi = \Phi_{\star} + \Phi_{\text{p}} + \Phi_{\text{ind}}$ is the net gravitational potential composed of terms from the star, the planet, and the indirect planet-star gravitational interactions. Here $\Phi_{\star} = -G\mstar/r$, $\Phi_{\text{p}}$ and $\Phi_{\text{ind}}$ are defined in Section \ref{sec:planet}. $\epsilon = \rhod / \rhog$ is the local dust-to-gas ratio. $\mathcal{T}$ is the viscous stress tensor given by 
\begin{equation}
\mathcal{T} = \rhog \nu \left[ \nabla \V + (\nabla \V\,)^{\dagger} - \frac{2}{3} \I\, \nabla \cdot \V\, \right],
\end{equation}
where $\nu$ is the gas kinematic viscosity, and $\I$ is the identity tensor. 

We adopt a constant $\nu = 10^{-5} R_\text{ref}^2 \, \Omega_\text{K,ref}$, corresponding to $\alpha = 4 \times 10^{-3}$ at $R = R_\text{ref}$ in the conventional $\alpha$-viscosity prescription \citep{shakura_reprint_1973}. This is to suppress vertical shear instability \citep[VSI;][]{nelson_linear_2013}, which would otherwise stir up dust grains \citep{flock_radiation_2017, flock_gas_2020, lin_dust_2019}. Similarly, we intentionally omit dust diffusion, which is typically used to represent particle stirring by gas turbulence \citep[e.g.,][]{weber_predicting_2019}. The chosen viscosity value also suppresses vortex formation \citep{koller_vortices_2003, li_potential_2005, li_type_2009, lin_type_2010} at gap edges, which would introduce non-axisymmetry and may also interfere with dust settling \citep{zhu_particle_2014}. 

Our disk models are thus designed to minimize known mechanisms that hinder dust settling, so that we can focus on the influence of planet-induced gas flows on axisymmetric dust rings. 


\subsection{Planet} \label{sec:planet}

We consider a planet on a fixed, circular orbit at $R = R_\text{ref}$ on the disk midplane. Our fiducial planet mass is $M_{\text{p}} = 3\times10^{-4}\mstar$, corresponding to a Saturn-mass planet around a solar-mass star, which is sufficient to open a gas gap \citep{kanagawa_mass_2016}. The planet-related potential terms are
\begin{equation}
\Phi_{\text{p}} + \Phi_{\text{ind}} = - \frac{Gm_{\text{p}}(t)}{\sqrt{r^{\prime^2} + r_{\text{s}}^2}} + \frac{Gm_{\text{p}}(t)}{R_\text{ref}^2}R\cos{(\phi - \phi_{\text{p}})},
\end{equation}
where $\phi_{\text{p}}$ is the azimuth of the planet, $r_{\text{s}} = 0.1H_{\text{g}}$ is a smoothing length,
\begin{equation}
r^\prime = \sqrt{R^2 + R_\text{ref}^2 - 2RR_\text{ref}\cos{(\phi - \phi_{\text{p}})} + Z^2}
\end{equation}
is the distance to the planet, and $m_\text{p}(t)$ is the time-dependent planet mass to avoid transient effects associated with the suddenly introduced massive planet. We switch on the planet's potential over a timescale $t_{\text{g}}$ by prescribing 
\begin{equation}
m_{\text{p}}(t) = 
\begin{cases}
0 
    & t = 0, \\
\frac{1}{2}\left[1 - \cos\left(\frac{t}{t_{\text{g}}}\pi\right)\right]M_{\text{p}} 
    & 0 < t < t_{\text{g}}, \\
M_{\text{p}} 
    & t \geq t_{\text{g}}.
\end{cases} 
\end{equation}
We adopt $t_{\text{g}} = 500 P_\text{ref}$, where $P_\text{ref} = 2\pi \Omega_\text{K,ref}^{-1}$ is the planet's orbital period.


\subsection{Gas and Dust Initialization} 

The gas density is initialized to 
\begin{equation} \label{eq:rhog}
\rho_\text{g0} = \rho_\text{g0,ref}\left(\frac{R}{R_\text{ref}}\right)^{-p} 
\times \exp\left[\frac{G\mstar}{\cs^2}\left(\frac{1}{r} - \frac{1}{R}\right)\right],
\end{equation}
with $p=1.5$. The initial reference midplane gas density $\rho_\text{g0,ref}$ is arbitrary for a non-self-gravitating disk. For a thin disk ($|Z| \ll R$) the vertical gas profile is Gaussian ($\propto \exp{[-Z^2/2H_{\text{g}}^2]}$). 

The dust density is initialized by $\rho_\text{d0} = \epsilon_0 \rho_\text{g0}$. The local dust-to-gas ratio $\epsilon$ is initialized to
\begin{equation}
\epsilon_0(R, Z) = \epsilon_{\text{0,mid}}(R) \times \exp{\left(-\frac{Z^2}{2H_{\epsilon}^2}\right)},
\end{equation}
where $\epsilon_\text{0,mid} = 0.1$ is the uniform initial midplane dust-to-gas ratio, except being tapered to zero at the radial boundaries. $H_{\epsilon}$ is defined as 
\begin{equation}
H_{\epsilon} = \frac{H_{\text{g}} H_{\text{d}}}{(H_{\text{g}}^2 - H_{\text{d}}^2)^{1/2}},
\end{equation}
such that the $\rho_\text{d0}$ profile remains vertically Gaussian. The value of $H_{\epsilon}$ is chosen such that the initial dust scale-height satisfies $H_{\text{d0}} = 0.1H_{\text{g}}$. Therefore, the initial dust-to-gas mass ratio, or metallicity, is $\sim$ 0.01.

We follow \cite{kanagawa_effect_2017} and initialize the gas and dust azimuthal velocities to 
\begin{align}
V_{\phi 0} = & \; R \, \Omega_\mathrm{K}(R)\left(\sqrt{1-2\eta} + 
    \frac{\epsilon_0\eta}{\epsilon_0+1}\frac{1}{\text{St}^{\prime2}+1}\right) \\
W_{\phi 0} = & \; r \, \Omega_\mathrm{K}(r) - 
    R \, \Omega_\mathrm{K}(R)\left(\frac{\eta}{\epsilon_0+1}\frac{1}{\text{St}^{\prime2}+1}\right),
\end{align}
where $\text{St}^{\prime} = \text{St}/(1+\epsilon)$, and
\begin{equation}
\eta = \frac{1}{2}\left[(p+q)h_{\text{g}}^2 + q \left(1 - \frac{R}{r}\right)\right]
\end{equation}
is a dimensionless measurement of the global radial pressure gradient. The radial velocities are initialized to
\begin{alignat}{2}
V_{\text{R0}} = & &&\frac{2\epsilon_0\eta}{\epsilon_0+1}
    \frac{\text{St}^{\prime}}{\text{St}^{\prime2}+1}R\,\Omega_\text{K}(R) \\
W_{\text{R0}} = & - &&\frac{2\eta}{\epsilon_0+1}
    \frac{\text{St}^{\prime}}{\text{St}^{\prime2}+1}R\,\Omega_\text{K}(R).
\end{alignat}
These correspond to the inward drift of dust due to the radial pressure gradient and a compensating outward drift of gas due to angular momentum conservation. The initial vertical velocities $V_{\text{Z0}}$ and $W_{\text{Z0}}$ are set to zero. 

Note that we neglect the viscous accretion \citep{lynden-bell_evolution_1974} in the initial gas velocity field. In a 2D disk, this viscous radial gas flow is given by 
\begin{equation}
V_{\text{vis}}^{\text{2D}} = -\frac{3\nu}{R} \frac{d\ln{(\nu\sigmag R^{1/2})}}{d\ln{R}},
\end{equation}
where $\sigmag$ is the gas surface density. Since we have $\sigmag \propto R^{-1/2}$ for our prescribed $\rhog$ profile in Equation \ref{eq:rhog}, $V_{\text{vis}}^{\text{2D}} = 0$ for a constant $\nu$ as considered throughout this work. Thus our disk models have no net viscous accretion in an averaged sense. 


\subsection{Numerical method}
 
We evolve the above dusty disk using the \textsc{fargo3d} code \citep{benitez-llambay_fargo3d_2016}. \textsc{fargo3d} is a general-purpose finite-difference code, and is particularly suited for simulating protoplanetary disks as it includes the FARGO algorithm \citep{masset_fargo_2000} that alleviates time-step constraints imposed by the fast rotation at the inner disk boundary. 

We adopt a spherical domain centered on the star with $r \in [0.2, 4.0] \,R_\text{ref}$, $\phi \in [0, 2\pi]$, and polar angle such that $\tan(\frac{1}{2}\pi - \theta) \in [-3, 3] \,h_{\text{g}}$ (i.e., three gas scale-heights above/below the midplane). The resolutions we choose are $N_{r} \times N_\theta \times N_\phi = 360 \times 90 \times 720$, with logarithmic spacing in \textit{r} and uniform spacing in $\theta$ and $\phi$. We thus resolve $H_{\text{g}}$ by approximately 15 cells vertically, and 6 cells radially and azimuthally. The simulations are performed in the co-rotating frame with the planet.

The gas density is damped to its initial value at the radial boundaries, and is assumed to be in vertical hydrostatic equilibrium at the vertical boundaries. The dust density is symmetric at both radial and vertical boundaries. The meridional velocities of gas and dust are set to zero at the radial and vertical boundaries, except that the inner radial boundary is open for mass loss of dust. The azimuthal velocities at those boundaries are assigned at the Keplerian speed with a pressure offset for gas. Periodic boundaries are imposed in the $\phi$ direction.


\begin{figure*}[t]
\includegraphics[width = 1.0\textwidth]{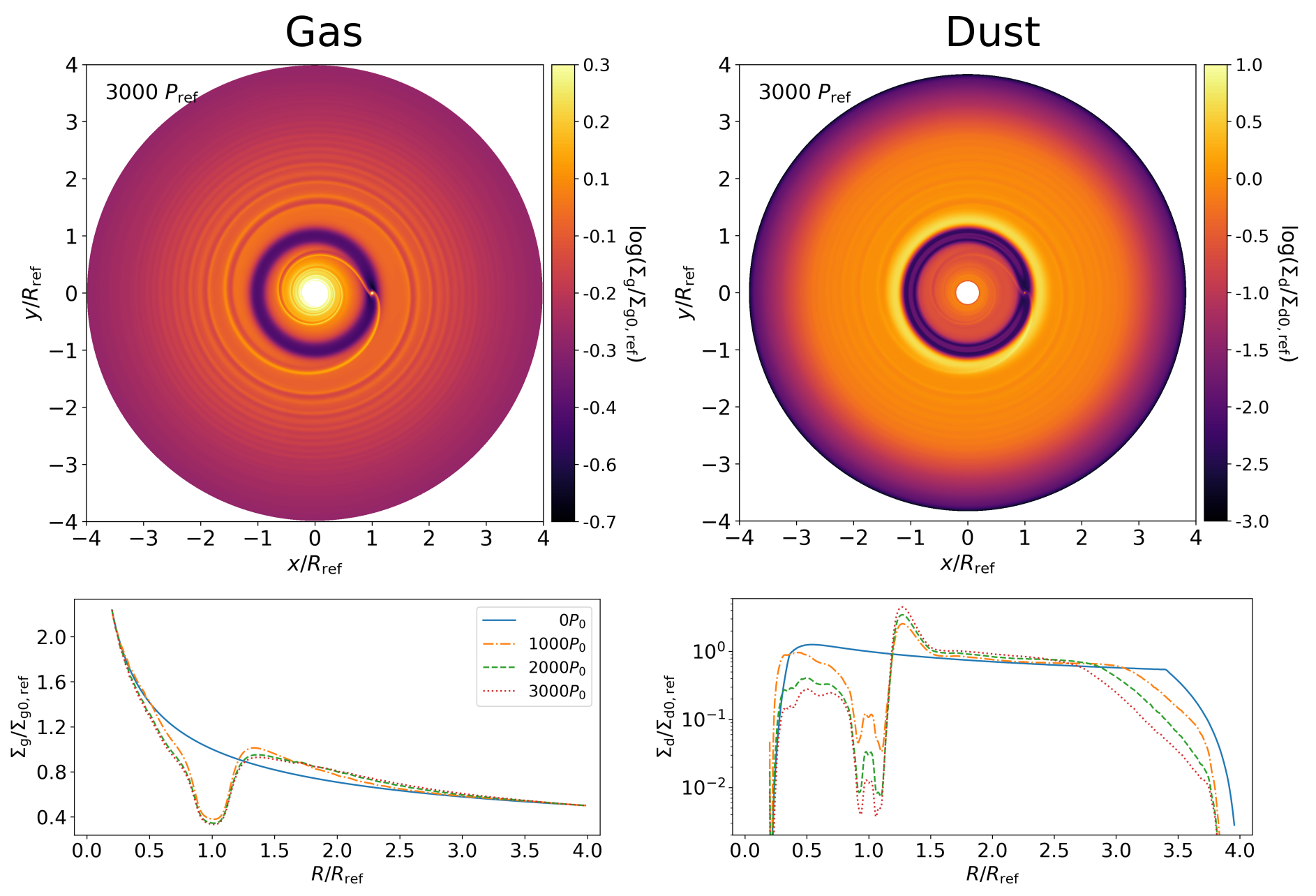}
\figcaption{
Distributions of gas (left) and dust (right) surface density from the fiducial simulation with a Saturn-mass planet and dust grains with $\text{St}_\text{0,ref}=10^{-3}$. 
\textit{Upper panels}: Ecliptic plots at 3000$P_\text{ref}$. 
\textit{Lower panels}: Azimuthally averaged (excluding azimuth within $\arcsin(3R_{\text{H}}/R_\text{ref})$ to the planet) radial profiles at 0, 1000, 2000, and 3000$P_\text{ref}$.
All panels are normalized to the initial value at $R = R_\text{ref}$.
\label{fig:ecliptic}} 
\end{figure*}


\subsection{Diagnostics}

To analyze our results, we first convert simulation outputs from spherical to cylindrical coordinates using cubic interpolations. Motivated by observations of rings and gaps, we mostly examine azimuthally averaged profiles, which is representative since non-axisymmetric dust features are weak and only appear close to the planet. Nevertheless, we mask the planet by omitting values in the region $\phi - \phi_{\text{p}} < \psi$ when calculating azimuthal averages. $\psi = \arcsin{(3R_{\text{H}}/R_\text{ref})}$ denotes the angular distance of three Hill radii, with $R_{\text{H}} = R_\text{ref}\sqrt[3]{M_{\text{p}}/(3\mstar)}$. 

We are primarily interested in the thickness of the dust layer. We define and obtain the dust scale-height $H_{\text{d}}(R)$ by fitting 
\begin{equation}\label{hdust_gauss}
\rhod(R, Z) = \rho_\text{d,mid}(R) \times \exp\left({-\frac{Z^2}{2H_{\text{d}}^2}}\right).
\end{equation}
However, for well-settled dust layers (see Section \ref{sec:planetmass}) we find it impractical to accurately fit them with Gaussians\footnote{The least-squares fitting function returns the input parameters of initial guess instead of solutions corresponding to a local minimum of the cost function.}. In such cases we obtain $H_{\text{d}}$ by searching where 
\begin{equation}
\rhod(R, 2H_{\text{d}}) = \rho_\text{d,mid}(R) \times \exp(-2)
\end{equation}
is satisfied. This definition of the dust layer thickness is more robust, and coincides with Equation \ref{hdust_gauss} when the distribution is close to Gaussian. 


\section{Results} \label{sec:result}

In this section, we first compare our fiducial result with previous 2D studies. Secondly, we describe the dust kinematics in our fiducial simulation and discuss the origin of it. Finally, we briefly explore the effect of disk parameters on the reported dust behavior.  


\subsection{Disk Morphology in the Ecliptic Plane} \label{sec:ecliptic}

Figure \ref{fig:ecliptic} shows the gas and dust surface density distributions at $3000 P_\text{ref}$. The planet opens a gap of 5--6 $R_{\text{H}}$ wide in the gas disk, which is consistent with the empirical result in \cite{kanagawa_mass_2016} and \cite{dong_what_2017} based on 2D simulations. A gas bump is formed on the outer gap edge due to gas being evacuated from the gap by the planet. However, its counterpart at the inner edge is smoothed out by the inward gas flows driven by the net effect of negative planetary and viscous torques. Dust is more cleared compared with gas, but there are still some remaining in the horseshoe orbits. We note the axisymmetric dust ring at the outer gap edge that indicates dust trapping on site. Other than the planet-induced spiral arms, non-axisymmetric features are weak in the disk. 

Although our models are 3D, these surface density maps are similar to early 2D simulations \citep{paardekooper_dust_2006}. More importantly, the dust gap of sub-mm-sized grains induced by our Saturn-mass planet, which is equivalent to $100\mearth$ or $0.3M_\text{J}$, is expected to be observable by ALMA \citep{rosotti_minimum_2016}. 

The similarity between 2D and 3D simulations in the gas surface densities has already been pointed out by \cite{fung_gap_2016}, which indicates that 2D simulations are sufficient to obtain a representative gas morphology. On the other hand, we find below that the vertical distribution of dust can be significantly affected by the planet, which requires 3D modeling. 


\subsection{Dust Stirred up above the Disk Midplane} \label{sec:puffup}

Figure \ref{fig:dgratio} shows the meridional snapshots of the local dust-to-gas ratio at ages 0, 500 (when the planet is fully introduced), and 3000$P_\text{ref}$, respectively. Figure \ref{fig:radprof} shows the time evolution of the normalized dust scale-height $H_{\text{d}}/H_{\text{g}}$, and the particle stopping time $\tau_\text{s}$ at different disk heights at 3000$P_\text{ref}$. Once the planet starts to open the gap, we find dust is efficiently stirred-up\footnote{Hereafter, the upward direction refers to the one pointing away from the midpane, and the downward direction refers to the one pointing toward the midplane.} above the disk midplane with a characteristic dust scale-height $H_\text{d} \sim 0.7H_\text{g}$. As is shown in the Appendix \ref{app:slices}, this dust `puff-up' phenomenon is nearly axisymmetric.

\begin{center}
\includegraphics[width = 0.45\textwidth]{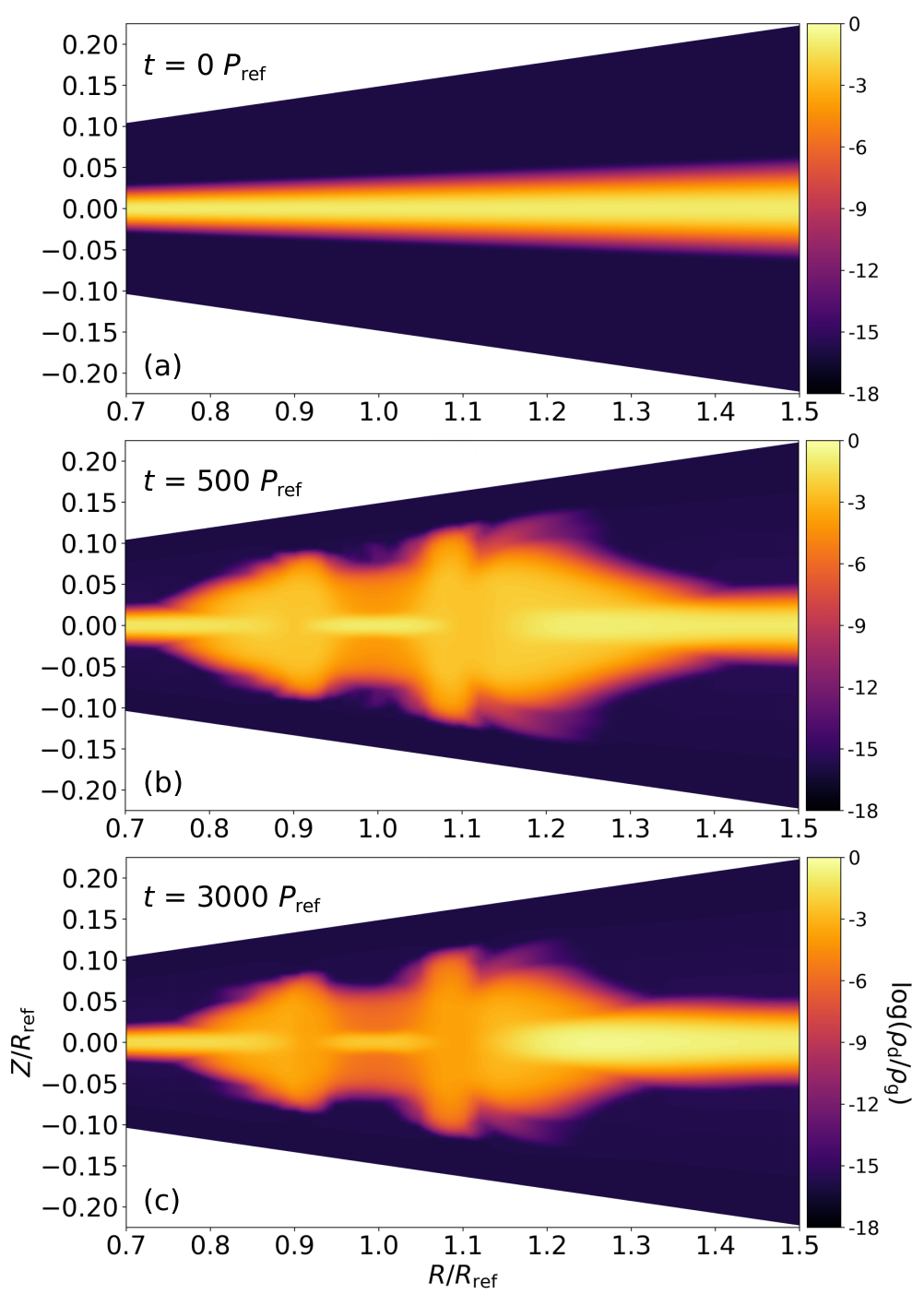}
\figcaption{
Meridional distributions of the azimuthally averaged (excluding azimuth within $\arcsin(3R_{\text{H}}/R_\text{ref})$ to the planet) dust-to-gas ratio at 0, 500 (when the planet potential is fully switched on)}, and 3000$P_\text{ref}$ from the fiducial simulation with a Saturn-mass planet and dust grains of $\text{St}_\text{0,ref} = 10^{-3}$. The planet is at $R=R_\text{ref}$. Colors are mapped in the logarithmic scale. 
\label{fig:dgratio}
\end{center}

Figure \ref{fig:stream} shows the streamlines of dust and gas plotted over the local dust-to-gas ratio at 3000$P_\text{ref}$. The flow pattern of gas qualitatively agrees with the results in \cite{fung_gap_2016}. It also shows that in the `puff-up' regions ($R \sim 0.9R_\text{ref}$ and $1.1R_\text{ref}$, $Z \lesssim H_{\text{g}}$), the dust and gas velocity streamlines are similar, indicating that the dust kinematics there is closely related to the gas kinematics.

\begin{center}
\includegraphics[width = 0.45\textwidth]{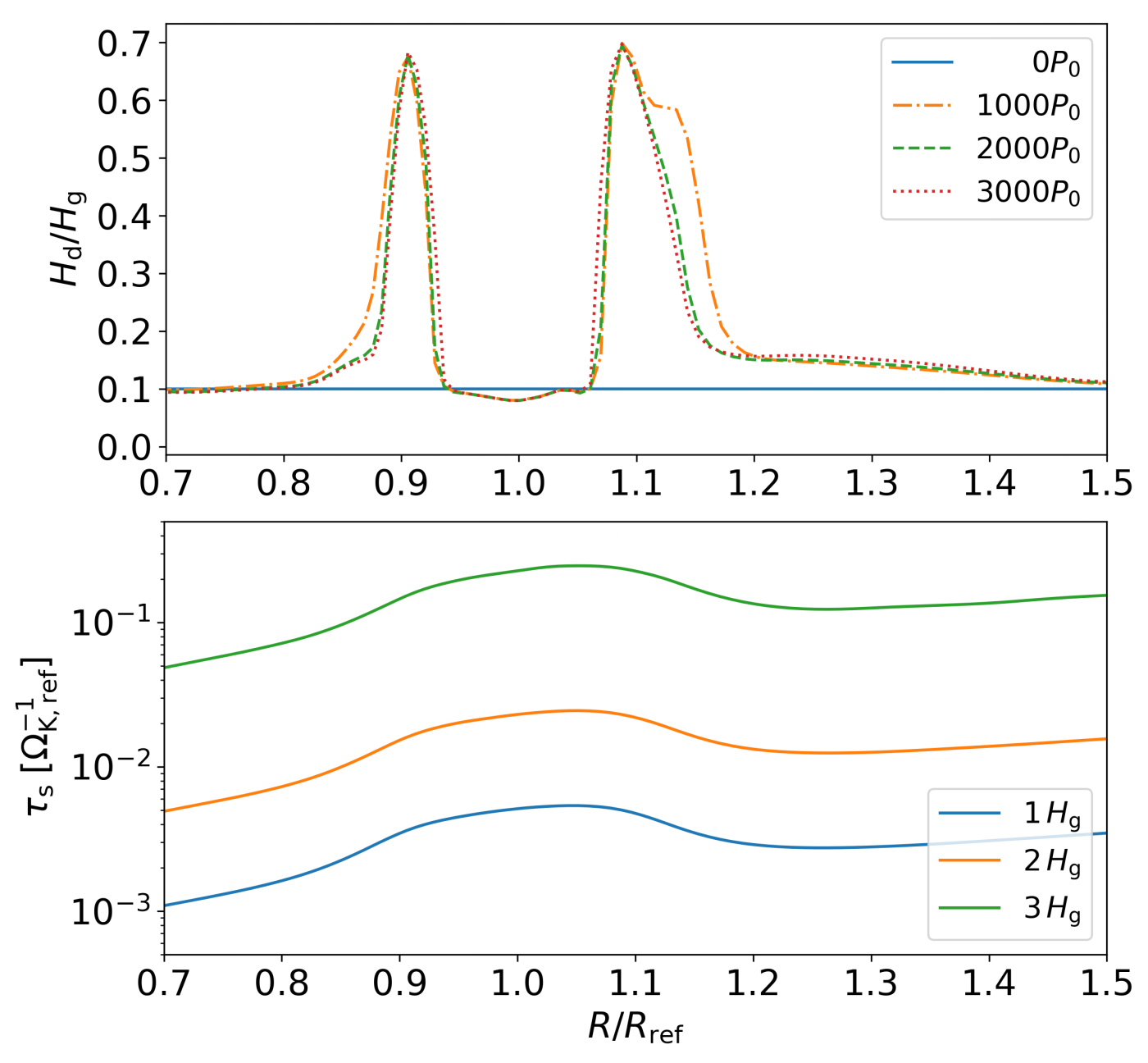}
\figcaption{
\textit{Upper:} Radial profiles of the normalized dust scale-height at 0, 1000, 2000, and 3000$P_\text{ref}$.
\textit{Lower:} Radial profiles of the particle stopping time $\tau_\text{s}$ at \textit{Z} = 1, 2, and 3 $H_{\text{g}}$ at 3000$P_\text{ref}$.
Both panels are azimuthally averaged (excluding azimuth within $\arcsin(3R_{\text{H}}/R_\text{ref})$ to the planet), and plotted from the fiducial simulation with a Saturn-mass planet and dust grains of $\text{St}_\text{0,ref} = 10^{-3}$. The planet is at $R=R_\text{ref}$.
\label{fig:radprof}}
\end{center}

This result is expected for the small grains we consider. In this limit, the gas and dust kinematics can be associated by the terminal velocity approximation \citep{youdin_streaming_2005, jacquet_linear_2011, price_fast_2015, lovascio_dynamics_2019}:
\begin{equation} \label{eq:approx}
\W = \V + \frac{\nabla P}{\rho}\tau_\text{s},
\end{equation} 
where $\rho = \rhog + \rhod$ is the total density. This approximation (validated in Appendix \ref{app:approx}) shows that, for a dust grain tightly coupled to the gas, its velocity is almost identical to the gas velocity, with a correction due to local pressure gradients. In laminar PPDs with no turbulence or planets, vertical hydrostatic equilbrium ($|V_{\text{Z}}| \sim 0$) implies that $(\nabla P)_Z$ approximately balances the vertical gravity from the star. The second term on the right-hand-side (RHS) thus leads to a downward $W_{\text{Z}}$ towards the disk midplane, i.e. dust settles \citep{dubrulle_dust_1995, takeuchi_radial_2002}. Thus, for dust to be stirred up, $V_{\text{Z}}$ must be directed away from the midplane and overwhelm vertical stellar gravity. 

\begin{center}
\includegraphics[width = 0.45\textwidth]{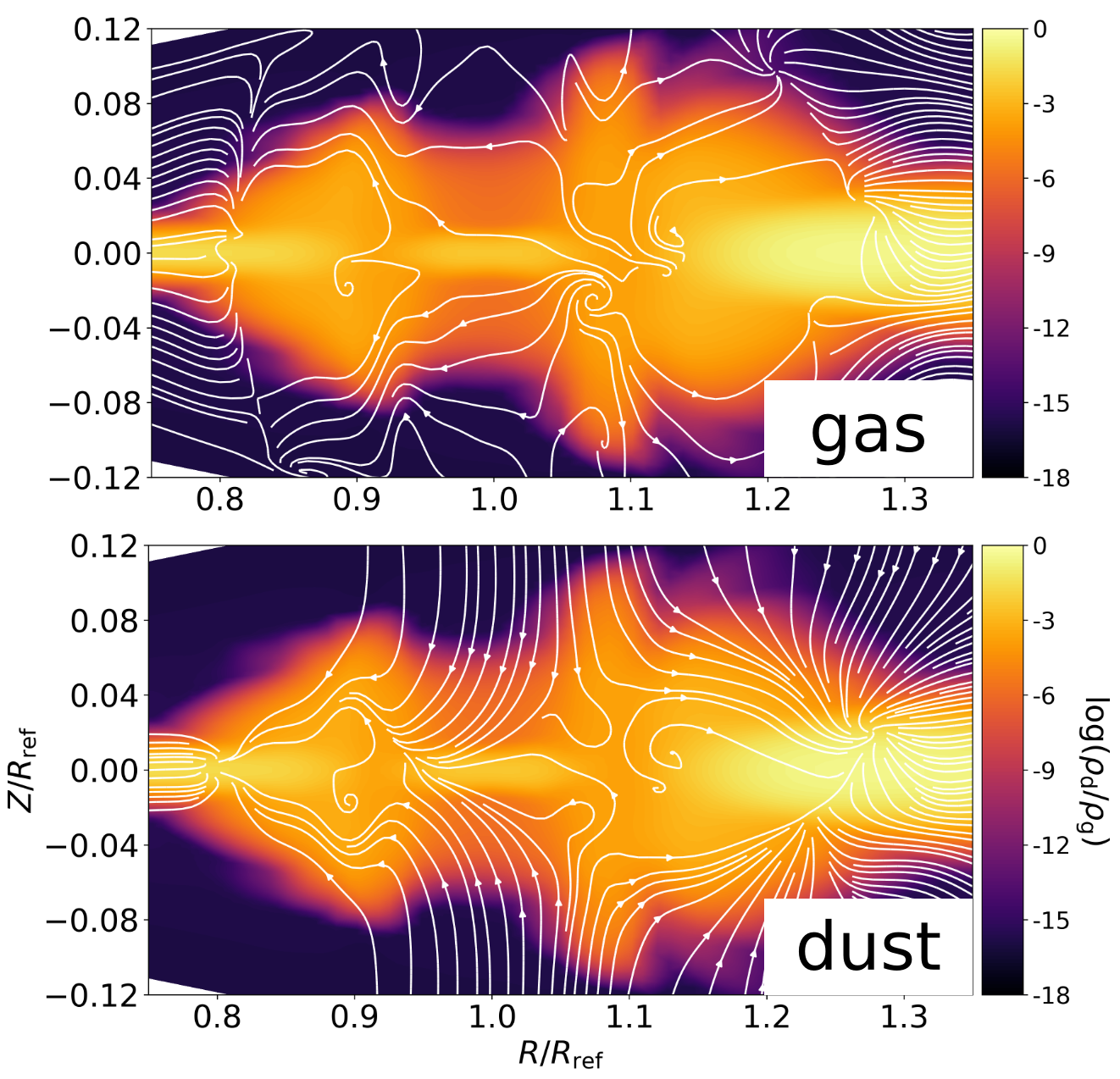}
\figcaption{
Streamlines of gas (upper) and dust (lower) flows at 3000$P_\text{ref}$ of the fiducial simulation with a Saturn-mass planet and dust grains of $\text{St}_\text{0,ref} = 10^{-3}$. Both panels are azimuthally averaged (excluding azimuth within $\arcsin(3R_{\text{H}}/R_\text{ref})$ to the planet), and plotted over the dust-to-gas ratio (part of Figure \ref{fig:dgratio}c). The planet is at $R=R_\text{ref}$. Both the vertical flows toward the planet and the flows repelled from the planet are masked out while averaging. 
\label{fig:stream}}
\end{center}

\begin{center}
\includegraphics[width = 0.45\textwidth]{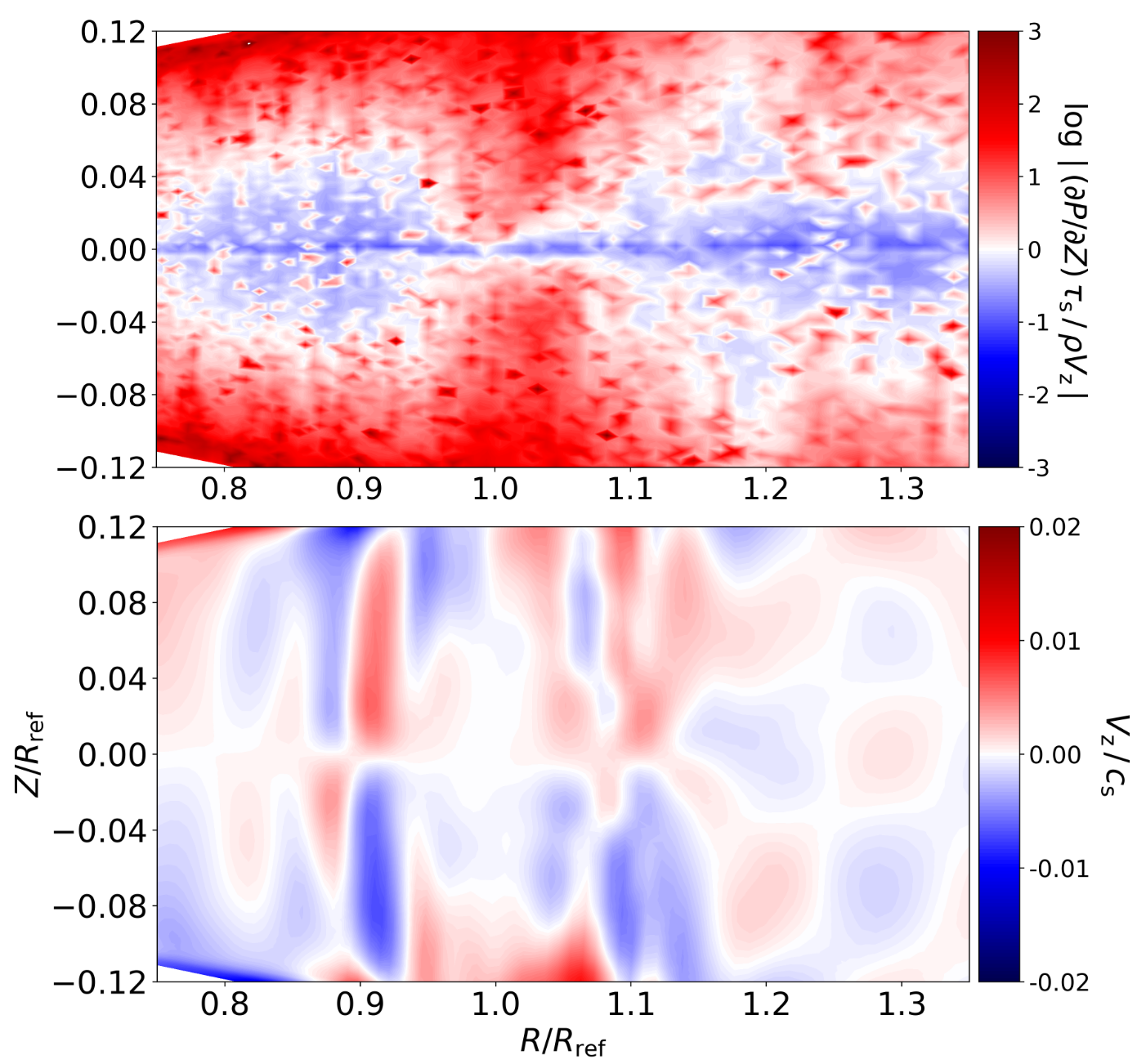}
\figcaption{
\textit{Upper}: The absolute ratio between the two terms on the right hand side of the terminal velocity approximation (Equation \ref{eq:approx}) in the vertical direction. Red (blue) indicates dust velocities are dominated by the local pressure gradient (gas flow). 
\textit{Lower}: The normalized gas vertical velocity. Red (blue) denotes positive (negative) velocities in \textit{Z} direction. 
Both panels are azimuthally averaged (excluding azimuth within $\arcsin(3R_{\text{H}}/R_\text{ref})$ to the planet), and taken at 3000$P_\text{ref}$ of the fiducial simulation (with a Saturn-mass planet and dust grains with $\text{St}_\text{0,ref}=10^{-3}$). The planet is at $R=R_\text{ref}$.
\label{fig:approxratio}} 
\end{center}

Figure \ref{fig:approxratio} shows the absolute ratio between the two terms on the RHS of Equation \ref{eq:approx}. A smaller ratio indicates a larger contribution to the dust velocity from the gas flow. We find that inside the gap, the vertical dust velocity is dominated by the pressure gradient throughout the vertical extent. Outside the gap, the same is true at high altitudes ($Z \gtrsim H_\text{g}$), leading to settling, while it is the other way around close to the midplane.

Figure \ref{fig:approxratio} also shows the normalized vertical gas velocity $V_{\text{Z}} / \cs$. At the `puff-up' radii ($\sim$ 0.9 and 1.1 $R_\text{ref}$), the upward gas velocity reaches $0.01 \cs$, roughly consistent with \cite{fung_gap_2016}. Given the dust settling speed $W_{\text{sed}} \sim \mathcal{O}(\cs \text{St})$ \citep{takeuchi_radial_2002}, we can expect the settling of grains to be disrupted by upward gas flows when $V_{\text{Z}} / \cs \gtrsim$ St, which is easily satisfied in our fiducial case with $\text{St} \sim 10^{-3}$.

The mechanism of planet-induced meridional gas flows has been analyzed in detail by \cite{fung_gap_2016}. The planet's Lindblad torques drive gap-opening gas flows away from the planet, which encounter gap-closing gas flows toward the planet driven by viscous torques. The magnitude of these opposing torques depend on height, and the net effect is the meridional circulation: an upward combined flow near gap edges and a downward flow close to the planet's orbit. Our simulations show that dust particles are then stirred up by these circulations.

Therefore, we conclude that the `puff-up' of dust at gap edges is due to planet-induced meridional gas flows.


\subsection{Effect of Stokes Number}

A larger grain size or Stokes number leads to weaker coupling between gas and dust. That is, the magnitude of the second term in the terminal velocity approximation increases with St (Equation \ref{eq:approx}), which favors dust settling. As discussed above, for Saturn-mass planets that induce vertical flows with $|V_{\text{Z}}| / \cs \sim 0.01$, we expect particles with $\text{St} \gtrsim 0.01$ to be much more resilient to being puffed up. We thus consider two simulations with $\text{St}_\text{0,ref} = 0.01$ and $\text{St}_\text{0,ref} = 0.1$, respectively. They are shown in Figure \ref{fig:stokes}b--d, which can be compared to the fiducial run in Figure \ref{fig:stokes}a.

We find that in the $\text{St}_\text{0,ref} = 0.01$ case (Figure \ref{fig:stokes}b), the `puff-up' is much weaker than the fiducial run at 3000$P_\text{ref}$, with $H_{\text{d}}/H_{\text{g}} \lesssim 0.2$ (cf., $\sim 0.7$ in the fiducial case). Due to the larger Stokes number, particles also drift inward more rapidly (Equation \ref{eq:approx}). We find a wide ($\sim 4H_\text{g}$) dust ring forms exterior to the planet due to a local pressure maximum at $1.3R_\text{ref}$, but it remains flat with $H_{\text{d}}/H_{\text{g}} \lesssim 0.15$, since planet-induced vertical flows of gas mostly lie closer in.

\begin{center}
\includegraphics[width = 0.45\textwidth]{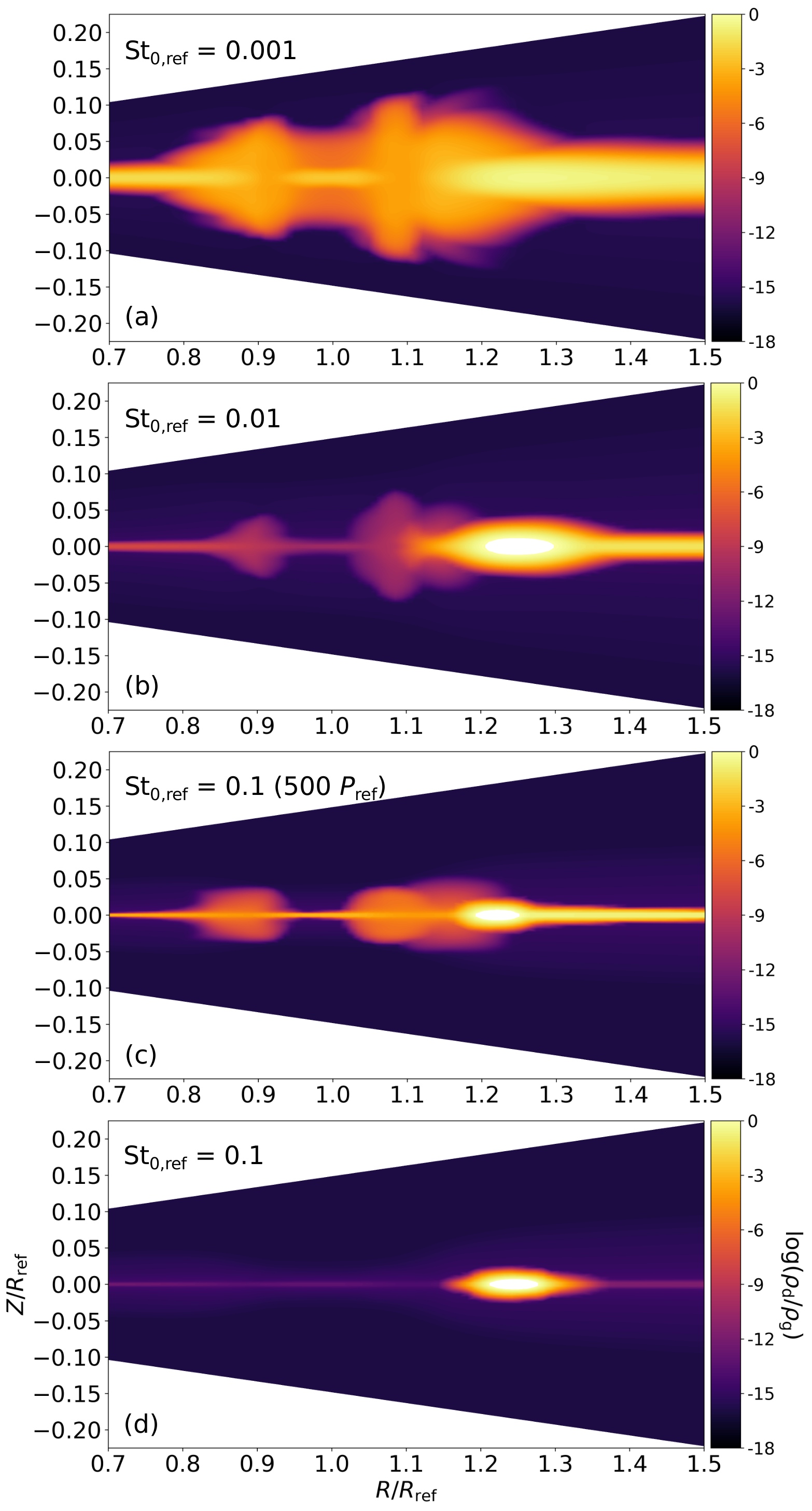}
\figcaption{
Meridional distributions of the azimuthally averaged (excluding azimuth within $\arcsin(3R_{\text{H}}/R_\text{ref})$ to the planet) dust-to-gas ratio around the planet's orbit at $R=R_\text{ref}$. (a): From the fiducial run (St$_\text{0,ref}$ = 0.001). (b): From the St$_\text{0,ref}$ = 0.01 run. (c)--(d): From the St$_\text{0,ref}$ = 0.1 run. Panel (a), (b) and (d) are taken at 3000$P_\text{ref}$, and (c) is taken at 500$P_\text{ref}$. 
\label{fig:stokes}}
\end{center}

For $\text{St}_\text{0,ref} = 0.1$, we find that planet-disk interactions can still trigger the dust `puff-up' (Figure \ref{fig:stokes}c), but this effect is transient as it disappears by 3000$P_\text{ref}$ (Figure \ref{fig:stokes}d). In this case, dust is rapidly lost through radial drift, leaving only dust trapped at the local pressure maximum exterior to the planet. The end result is a single, well-settled dust ring with thickness $H_{\text{d}}/H_{\text{g}} \lesssim 0.05$ and width $\sim 2H_{\text{g}}$ at $1.25R_\text{ref}$. 


\begin{center}
\includegraphics[width = 0.45\textwidth]{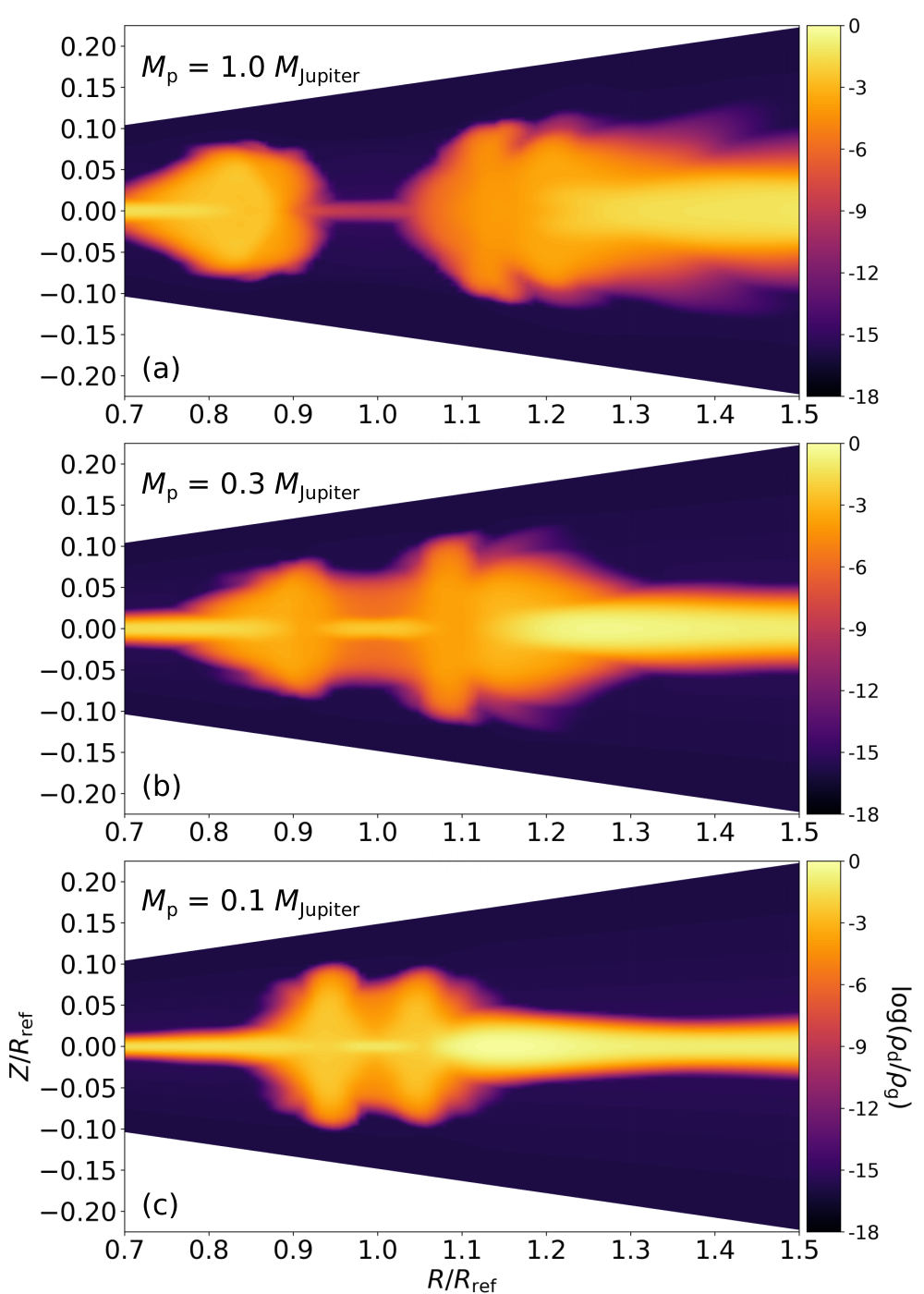}
\figcaption{
Meridional distributions of the azimuthally averaged (excluding azimuth within $\arcsin(3R_{\text{H}}/R_\text{ref})$ to the planet) dust-to-gas ratio at $3000 P_\text{ref}$ around the planet's orbit at $R=R_\text{ref}$. (a): From the Jupiter-mass planet run. (b): From the fiducial run (Saturn-mass/0.3 Jupiter-mass planet). (c): From the 0.1 Jupiter-mass planet run.
\label{fig:planet}}
\end{center}

\subsection{Effect of Planet Mass} \label{sec:planetmass}

A more massive planet is expected not only to create a deeper and wider gap \citep{kanagawa_mass_2016, fung_how_2014}, but also to drive stronger meridional gas flows around gap edges \citep{fung_gap_2016}. We here repeat the fiducial run with a Jupiter-mass planet ($M_\text{p} = 10^{-3}\mstar$). The resulting dust-to-gas ratio distribution at 3000$P_\text{ref}$ shown in Figure \ref{fig:planet}a.

We find that a higher planet mass leads to stronger and more complicated meridional gas flows around gap edges, with a typical upward gas velocity of $V_{\text{Z}} / \cs \sim 0.02$. As a result, a higher dust scale-height $H_{\text{d}} \sim 0.8H_{\text{g}}$ is found at the inner gap edge. However, a dust scale-height $H_{\text{d}} \sim 0.6H_{\text{g}}$ lower than the fiducial is found at the outer gap edge. We attribute this effect to a larger particle stopping time at the outer gap edge due to stronger gap-opening effect by the Jupiter-mass planet. This produces a lower gas density that lessens the dust-gas coupling (since $\tau_\text{s}\propto \rhog^{-1}$, see Equation \ref{eq:tstop}) and leads to more efficient dust settling at the outer gap edge.

To test if sub-mm-sized grains can settle against a less massive, but still gap-opening planet, we also present a simulation with $M_{\text{p}} = 10^{-4} \mstar$, corresponding to $0.1 M_{\text{J}}$ or $30 \mearth$. For comparison, \cite{rosotti_minimum_2016} find $M_{\text{p}} \gtrsim 20 \mearth$ is needed to induce dust rings \citep[see also][]{lambrechts_separating_2014}. The result is shown in Figure \ref{fig:planet}c, where we find that the dust `puff-up' still exists, with $H_{\text{d}}/H_{\text{g}}$ at the inner and outer gap edge being $\sim$ 0.5 and 0.4, respectively.


\subsection{Effect of Metallicity}

Our simulations include full back-reaction from dust onto gas. In this case, the global dust-to-gas ratio, or metallicity, can be important when considering vertical flows. This is because metallicity introduces a stabilizing effective buoyancy force \citep{lin_thermodynamic_2017}, which can favor dust settling \citep{lin_dust_2019} and may potentially resist dust puff up.

We repeated the fiducial run with a midplane dust-to-gas ratio $\epsilon_\text{0,mid} = 1$ (so the global metallicity is $\sim 0.1$). Dust back-reaction is expected to be non-negligible in this case. We found that the dust `puff-up' is quite similar to that in the fiducial case. Thus the dust back-reaction is unimportant to the `puff-up'. This is likely due to the small Stokes numbers considered herein, so that the dust-gas system behaves close to a single fluid and both are stirred up by the planet. 


\section{Discussion} \label{sec:discuss}

Here we briefly discuss a few inspirations that the planet-induced dust `puff-up' effect has brought, and introduce a few aspects that can be fulfilled in the future studies.

\subsection{Dust settling against planet-stirring}

The finite thickness of dust layers is usually attributed to particle stirring by gas turbulence \citep{fromang_dust_2006, zhu_dust_2015, flock_radiation_2017, flock_gas_2020}, which can be modeled as a diffusion process with the particle scale-height given by 
\begin{align}\label{eq:hdust_theory} 
    H_\text{d} = \sqrt{\frac{\delta}{\text{St} + \delta}}H_\text{g}
\end{align}
\citep{dubrulle_dust_1995, youdin_particle_2007}, where $\delta$ is a dimensionless measurement of particle diffusion by gas turbulence. Our simulations are laminar, and particles are stirred by planet-induced, vertically-global meridional flows \citep{fung_gap_2016} instead of turbulence. Nevertheless, we could apply Equation \ref{eq:hdust_theory} to our fiducial simulation to obtain an effective $\delta \sim \mathcal{O}(10^{-3})$ at gap edges, to place our results in the wider context of dust settling in protoplanetary disks.


\subsection{Inspirations to Observations}

Recent disk surveys have shown that dust gaps and rings are common in PPDs \citep{andrews_disk_2018, huang_disk_2018, long_gaps_2018, van_der_marel_protoplanetary_2019}. The origin of these rings are still being debated, although disk-planet interaction is frequently invoked \citep{dipierro_planet_2015, liu_new_2018, muley_pds_2019, toci_long-lived_2020, pinte_nine_2020}. Here, a common approach is to first perform 2D disk-planet simulations, then construct a 3D dust distribution assuming a turbulent diffusion model \citep[e.g., Equation \ref{eq:hdust_theory};][]{dubrulle_dust_1995}, before producing a synthetic image via radiative transfer and comparing it with observations  \citep[e.g.,][]{dong_observational_2015-1, jin_modeling_2016, facchini_annular_2020}. 

However, our simulations show that while gas-gap-opening planets naturally produce dust rings, they also easily stir up sub-mm-sized grains to high elevations. The vertical dust distribution is non-trivial. It must therefore either be obtained from explicit 3D disk-planet simulations, or otherwise accounted for using knowledge synthesized from 3D simulations. Dust structures with realistic vertical extent should be used as input for radiative transfer calculations to produce synthetic images, especially when the sharpness of the rings and gaps are concerned.

Observations, on the other hand, indicate that at least some observed dust rings are well-settled due to their sharp appearance \citep{pinte_dust_2016}. We thus suggest that the observed or inferred dust layer thickness may be used to distinguish dust rings formed by gas-gap-opening planets from those formed by other mechanisms that do not simultaneously stir up dust \citep[e.g., snowlines,][]{zhang_evidence_2015}. That is, the former should produce thicker dust rings. Conversely, within the planet interpretation, the sharpness of dust rings can be used to place an lower limit on the grain size. 

For example, \cite{jin_modeling_2016} find that three planets of around Saturn mass can account for the three well-defined dust gaps observed around HL Tau, assuming mm-sized grains. In their disk models such grains have Stokes numbers around $10^{-3}$ to $10^{-2}$ around the gaps. These parameter values are comparable to our fiducial case. We thus expect dust rings (or gap edges) to be puffed-up, making the gaps and rings less prominent at modest inclinations. This suggests grains should in fact be larger than mm-sized. 

There are observational evidence for puffed-up dust rings. \cite{doi_estimate_2021} revisited the ALMA observation of the HD 163296 disk \citep{andrews_disk_2018}, and found  $H_\text{d}/H_\text{g}>0.57$ at the B67 ring and $H_\text{d}/H_\text{g}<0.4$ at the B100 ring, assuming they are optically thin. \cite{huang_multifrequency_2020} found that a dust ring at $\sim$ 84 au in the disk around GM Aur has different shapes between its inner and outer edge. One possible explanation is that the ring has a finite thickness. Considering dust grains of 0.1 mm size at that location, we find a Stokes number of $\sim 5\times10^{-3}$, assuming a total disk mass of $0.18 \;\msun$, an outer disk radius of 450 au, a surface density power-law index of -1.5, and a grain density of $1.5 \;\text{g cm}^{-3}$ \citep{mcclure_mass_2016, huang_multifrequency_2020}. Our simulations then suggest that the observed ring morphology might result from dust stirred up by a young, Saturnian planet in the disk, which is within the planet mass range estimated by \cite{huang_multifrequency_2020} to carve the dust gap just interior to the dust ring.

Finally, although planets are one of the most popular explanations for rings and gaps, only a few planets have been directly detected in PPDs \citep[e.g., PDS 70b/c;][]{keppler_discovery_2018, haffert_two_2019}. Given that many of the disks are not face-on, we suggest that there could be planets obscured by the walls of the dust `puff-up', causing extinction of planets in observations of near infrared and optical wavelengths. 


\subsection{Caveats and outlooks}

The cost of full 3D simulations only allowed us to conduct a brief parameter study to qualitatively show that dust stirred up by gap-opening planets is a robust phenomenon. A larger number of simulations is needed in more comprehensive studies to work out an empirical formula for the dust `puff-up' as a function of grain size and planet mass. For example, one can fit the dust scale-height according to Equation \ref{eq:hdust_theory} and quantify the effective particle diffusion caused by planet-stirring as a function of planet mass and dust size.

We have considered planet masses larger than the thermal mass $M_\text{th} = h_{\text{g}}^3\mstar$ ($\sim 10^{-4}\mstar$), for which planets are expected to open gas gaps \citep{korycansky_method_1996, goodman_planetary_2001}. However, lower planet masses can open gaps in dust with little perturbation to the gas \citep{rosotti_minimum_2016, dipierro_opening_2017, dong_multiple_2017, chen_dusty_2018}. Such planets may not induce strong 3D meridional flows to stir up dust grains. Thus, low-mass planets may be consistent with well-defined and settled dust rings. Explicit simulations will be required to test this hypothesis. 

In this work we adopt a high kinematic viscosity $\nu = 10^{-5} R_\text{ref}^2 \Omega_\text{K,ref}$. This value of viscosity, and the neglect of a corresponding dust diffusion \citep{youdin_particle_2007}, were purposely chosen to eliminate dust-lofting mechanisms other than the planet-induced meridional gas flows. In real PPDs, the effective viscosity can be attributed to hydrodynamic turbulence, which provides additional particle stirring. For example, VSI can lead to weak turbulence \citep[$\alpha \sim 10^{-4}$ for $h_\text{g}=0.05$;][]{manger_high_2020}, but is sufficient to carry small dust to the atmosphere \citep{stoll_particle_2016}, which may lead to even more significant dust `puff-ups' if combined with planet stirring.

On the other hand, dust settling against VSI is sensitive to metallicity \citep[e.g.,][]{lin_dust_2019}, which is expected to increase around the outer gap edge as dust accumulates there. In order to study the combined effect of multiple mechanisms (e.g., VSI plus planet-stirring), low viscosity (or inviscid) simulations are needed in future work, which also require much higher resolutions to resolve small-scale turbulent motions \citep{picogna_particle_2018, manger_high_2020}.


\section{Conclusions} \label{sec:summary}

In this paper, we use 3D hydrodynamic simulations to study the dust kinematics in protoplanetary disks where a planet is present. Our main findings are: 

\begin{enumerate}

\item Small, sub-mm-sized dust grains relatively well-coupled to the gas can be carried to higher disk elevations by the meridional flows around edges of gas gaps caused by a gap-opening planet. In the case of a Saturn-mass planet, 0.1-mm-sized grains can be puffed up to achieve a vertical scale height $\sim70\%$ of the gas.

\item Grain size is the primary factor that affect the dust `puff-up'. Larger ($\gtrsim$ mm-sized) grains can settle against the stirring by a Saturn-mass planet. While large planet masses produce stronger `puff-ups', even weakly gas-gap-opening planets can stir up sub-mm-sized dust grains. For the Stokes number we consider in this study, the metallicity up to a global average of $\sim$0.1 is unimportant to the dust `puff-up'.

\end{enumerate}

We conclude that dust grains should exceed mm-size when attributing well-settled dust rings to gas-gap-opening planets. We thus caution that explicit 3D simulations cannot be replaced by 2D ones followed by a vertical expansion without taking into account the stirring effect by planet-induced meridional flows when studying the sharpness of the mm-wavelength emission associated with gas gaps opened by planets. However, our results do not rule out the possibility that low-mass planets (e.g., super-Earths) may open dust gaps without perturbing the gas significantly \citep[e.g.,][]{dong_multiple_2017}, thereby allowing dust rings to settle.


\bigbreak

We thank the anonymous referee for constructive suggestions that largely improved the quality of the paper. We also thank Logan Francis and Jeffrey Fung for useful discussions. This work was initiated at the 2019 ASIAA Summer Student Program.  Simulations were carried out on the TAIWANIA-2 GPU cluster hosted by the National Center for High-performance Computing.
M.-K.L. is supported by the Ministry of Science and Technology under grant 107-2112-M-001-043-MY3 and an Academia Sinica Career Development Award (AS-CDA-110-M06). J.B. and R.D. are supported by the Natural Sciences and Engineering Research Council of Canada. R.D. acknowledges support from the Alfred P. Sloan Foundation via a Sloan Research Fellowship.

\clearpage
\bibliography{jiaqing}


\clearpage
\appendix


\section{Azimuthal slices of the dust-to-gas ratio and the gas vertical velocity}
\label{app:slices}

\begin{figure*}[h]
\begin{center}
\includegraphics[width = \textwidth]{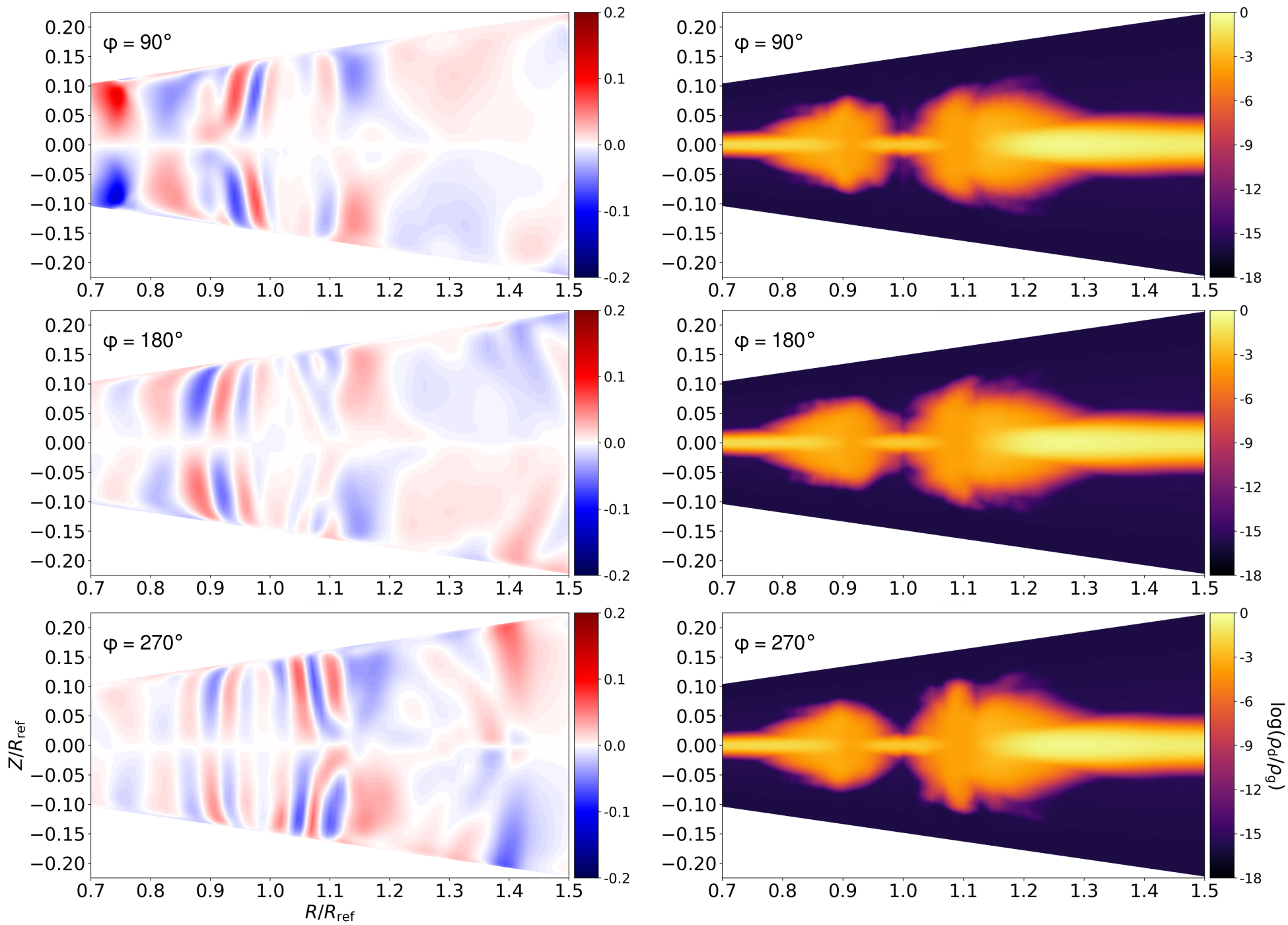}
\figcaption{Azimuthal slices of the gas vertical velocity (left) and the local dust-to-gas ratio (right) at $90^\circ$ (top), $180^\circ$ (middle), and $270^\circ$ (bottom) from the fiducial simulation with a Saturn-mass planet and dust grains of $\text{St}_\text{0,ref} = 10^{-3}$. The planet is at $R=R_\text{ref}$, $\phi = 0$.
\label{fig:slices}} 
\end{center}
\end{figure*}

In Section \ref{sec:puffup} we show azimuthally averaged plots and conclude that the origin of the dust `puff-up' phenomenon is the planet-induced gas meridional flows. However, azimuthally averaged plots cannot demonstrate any azimuthal behavior of the dust `puff-up'. Here in Figure \ref{fig:slices} we show azimuthal slices of the dust-to-gas ratio and the gas vertical velocity at $\phi = 90^\circ, 180^\circ, \text{and}\ 270^\circ$ from the fiducial simulation at 3000$P_\text{ref}$, to demonstrate that the dust `puff-up' is azimuthally global.

\clearpage

\section{Verification of the Terminal Velocity Approximation}
\label{app:approx}

\begin{figure*}[h]
\begin{center}
\includegraphics[width = 0.45\textwidth]{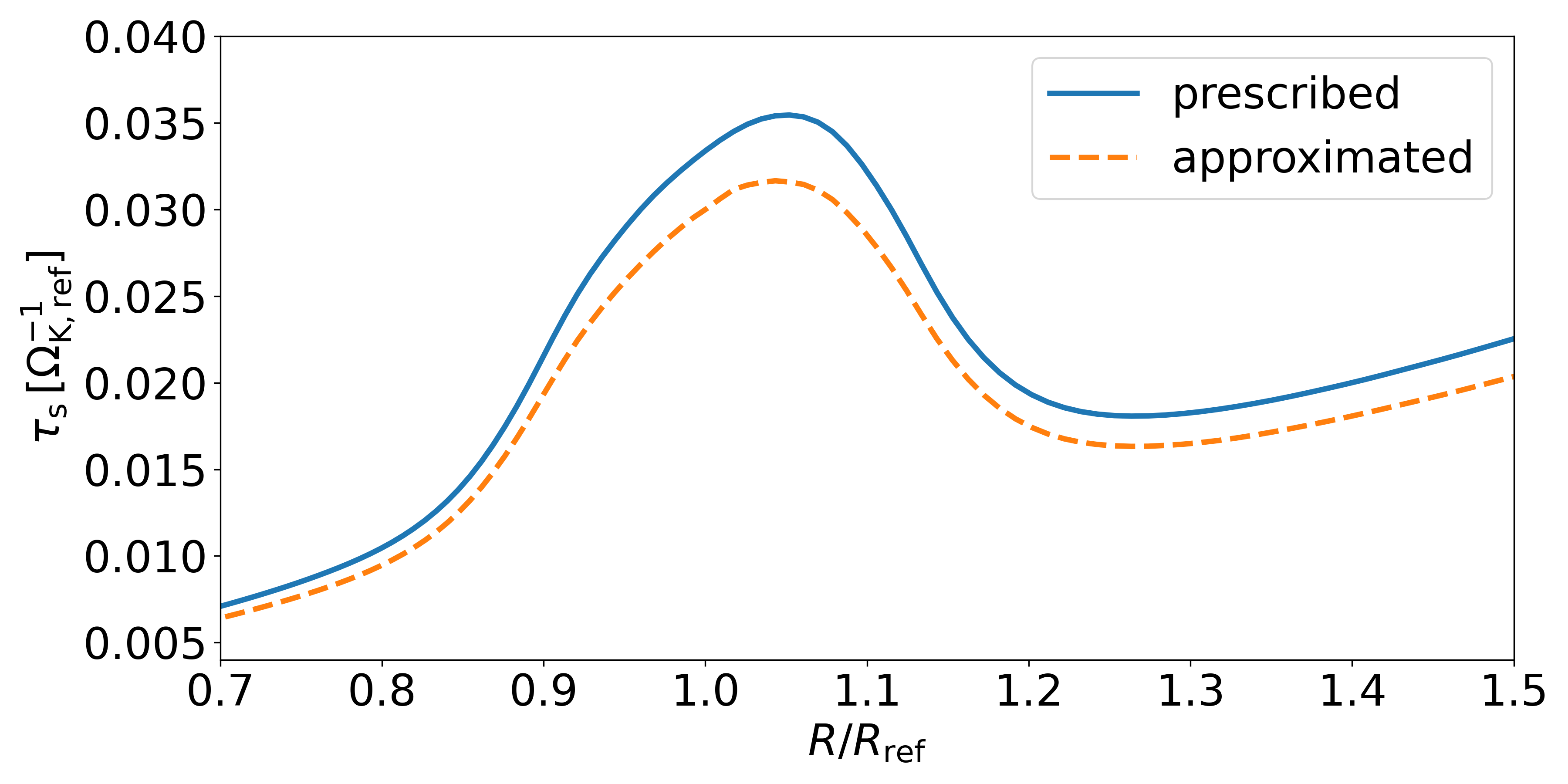}
\figcaption{
Radial profiles of the azimuthally and vertically averaged (excluding azimuth within $\arcsin(3R_{\text{H}}/R_\text{ref})$ to the planet) particle stopping time derived from Equation \ref{eq:tstop} (as prescribed in the simulations) and that  inferred from the terminal velocity approximation (Equation \ref{eq:approx}) in the vertical direction, respectively. The plot is taken at 3000$P_\text{ref}$ of the fiducial simulation with a Saturn-mass planet at $R=R_\text{ref}$ and dust grains with $\text{St}_\text{0,ref}=10^{-3}$. 
\label{fig:approx}} 
\end{center}
\end{figure*}

To validate the terminal velocity approximation used to interpret our simulations, we here compare $\tau_{\text{s}}$ derived from Equation \ref{eq:approx} in the vertical direction, namely $\tau_{\text{s}} = \rho(W_{\text{Z}} - V_{\text{Z}}) / (\partial P/\partial Z)$, with its prescribed definition in Equation \ref{eq:tstop}. Results for the fiducial simulation derived from the two approaches are shown in Figure \ref{fig:approx} after taking a vertical and azimuthal average. The largest relative difference between the two results is $\sim 10\%$, therefore we conclude that the dust kinematics in our study is well-described by the terminal velocity approximation as stated in Equation \ref{eq:approx}. 

\end{document}